\documentclass[aps,pra,twocolumn,amsmath,amssymb,nofootinbib,superscriptaddress]{revtex4-1}
\usepackage{times}
\usepackage[pdftex]{graphicx}
\usepackage{dcolumn}
\usepackage{bm}
\usepackage{amsmath}
\usepackage{indentfirst}
\usepackage{float}
\usepackage[colorlinks]{hyperref}
\usepackage[dvipsnames]{xcolor}

\usepackage{xcolor}

\newcommand{\adop}{\hat{a}^{\dagger}}

\newcommand{\bdop}{\hat{b}^{\dagger}}

\newcommand{\im}{{\rm i}}
\newcommand{\tr}{{\rm tr}}

\newcommand{\kron}{{\rm kron}}
\newcommand{\conj}{{\rm conjugate}}
\newcommand{\entropy}{S}

\newcommand{\svd}{{\rm svd}}
\newcommand{\perm}{{\rm perm}}
\newcommand{\detm}{{\rm det}}
\newcommand{\phaseshifter}{U_{{\rm ps}}}
\newcommand{\beamsplitter}{U_{{\rm bs}}}

\begin{document}

\title{Simulating the Dynamics of Single Photons in BosonSampling Devices with Matrix Product States}

\author{He-Liang Huang}
\author{Wan-Su Bao}
\author{Chu Guo} \email{guochu604b@gmail.com}

\affiliation{Henan Key Laboratory of Quantum Information and Cryptography, Zhengzhou, Henan 450000, China}
\affiliation{CAS Centre for Excellence and Synergetic Innovation Centre in Quantum Information and Quantum Physics,\\
University of Science and Technology of China, Hefei, Anhui 230026, China}
\date{\today}


\begin{abstract}
BosonSampling is a well-defined scheme for demonstrating quantum supremacy with photons in near term. Although relying only on multi-photon interference in nonadaptive linear-optical networks, it is hard to simulate classically. Here we study BosonSampling using matrix product states, a powerful method from quantum many-body physics. This method stores the instantaneous quantum state during the evolution of photons in the optical quantum circuit, which allows us to reveal the dynamical features of single photons in BosonSampling devices, such as entanglement entropy growth. We show the flexiblility of this method by also applying it to dissipative optical quantum circuits, as well as circuits with fermionic particles. Our work shows that matrix product states is a powerful platform to simulate optical quantum circuits. And it is readily extended to study quantum dynamics in multi-particle quantum walks beyond BosonSampling.

\end{abstract}

\maketitle

\section{Introduction}

Multi-photon interference lies at the heart of optical quantum information processing \cite{pan2012,sihui2012}, including optical quantum computing \cite{kok2007}, quantum cryptography \cite{lo2012}, quantum simulation \cite{Guzik2012}, and quantum metrology \cite{Giovannetti2011}. Based on some basic interference phenomena, such as the Hong-Ou-Mandel (HOM) effect \cite{hong1987}, it is possible for us to make use of quantum advantages to realize some complex processing tasks beyond the current classical limit \cite{Lund2017}, by increasing the number of single photons and the complexity of the optical quantum circuit.

In particular, BosonSampling \cite{Aaronson2013}, an intermediate model of linear optical quantum computing, utilizes only passive multi-port optical interferometer for the evolution of input single-photon sources, and then samples the output events using single-photon detectors, which is a well-defined problem that can be naturally and efficiently solved on a specialized photonic quantum simulator. However, it has been shown to be classically intractable because the amplitude associated with each term in the output quantum state of BosonSampling is related to the calculation of a matrix permanent \cite{Aaronson2013}, which is known to be a $\#{P}$-hard problem \cite{Valiant1979}. Thus, BosonSampling is considered as a promising candidate to experimentally demonstrate the $quantum$ $supremacy$ in the near future \cite{Harrow2017}. So far, a number of elegant BosonSampling experiments has been achieved with linear optics on a small scale \cite{Broome2013, Tillmann2013, Crespi2013, Spring2013, Spagnolo2014,Carolan2014,Carolan2015,he2017,wang2017,Loredo2017,Loredo2017,wang2018}. In addition, BosonSampling can be regarded as a special multi-particle quantum walk, and thus investigating its nonclassical correlations during the quantum dynamics is also an interesting task \cite{Urbina2014,Tamma2015,Laibacher2018,Shchesnovich2016,Walschaers2016,Huang2017}.

Previous works which use classical algorithms to solve the sampling task for BosonSampling have mostly focused on pushing the boundaries of quantum supremacy, namely to explore what is the largest scale that is reachable by the best classical algorithm and the best supercomputer in the world. In \cite{Neville2017}, it is reported that classical simulation of outputting one sample of a $30$-photon BosonSampling takes half an hour on a laptop, using the Metropolised independence sampling algorithm (MIS). In \cite{WuYang2018}, the permanent of a matrix with size $48\times48$ was calculated using Ryser's algorithm \cite{Ryser1963}, which takes $196608$ cores and an execution time of around $4500$ seconds on Tianhe-2 supercomputer. Recent work also show that BosonSampling does not necessarily require to compute a huge number of permanents to sample from the output distribution \cite{CliffordClifford2017}.

In this work, we focus on simulating the evolution of photons in the optical quantum circuit for BosonSampling (or, more broadly, the dynamical behavior of multi-particle quantum walks), rather than simulating the sampling tasks like those in the previous works. By simulating the behavior of photon in complex linear optical quantum circuits, it is possible to measure intermediate observables such as correlations, entanglement entropy, to have a better understanding of the underlying dynamical process, which is important for the study of multi-particle quantum walk. However, it is difficult to store the entire information about the evolution of multi-particle in a optical quantum circuit with $M$-mode and $N$-input photons. Firstly, considering the size of the Hilbert space it can easily be seen that, the number of parameters {\scriptsize{$\left( {\begin{array}{*{20}{c}}
{M + N - 1}\\
{M - 1}
\end{array}} \right)$}} of the output quantum state at each depth of the optical quantum circuit grows exponentially with the number of photons and modes in the system, thus one would soon run out of memory if one represents such a quantum state exactly on a classical computer. Secondly, for some special cases, such as BosonSampling, calculating a single amplitude of the output quantum state requires $O(2^nn^2)$ runtime according to Ryser's algorithm \cite{Ryser1963}.

We use matrix product states (MPS) method with a global U(1) symmetry (particle number conservation symmetry) to directly simulate the evolution of photons in the optical quantum circuit for BosonSampling. MPS has been a very successful numerical tool in solving one-dimensional quantum many-body problems \cite{Perez-GarciaCirac2007, Schollwock2011}, for both the unitary systems \cite{White1992, White1993, DelayVidal2004} and the dissipative systems \cite{GuoPoletti2018, BernierPoletti2018, XuPoletti2018}. Briefly speaking, MPS works by compactly rewriting a one-dimensional quantum manybody state as a product of many low dimensional tensors. For certain types of quantum systems which obey the so-called area law \cite{Hastings2007}, the sizes of the low dimensional tensors are bounded, which makes MPS to be a very memory efficient representation of the quantum state. The time complexity of MPS is also low in that the local unitary operations can be performed on local tensors only, without affecting other tensors. Moreover, measuring observables, especially local observables with MPS can be performed extremely efficiently \cite{Schollwock2011}.

For quantum systems with particle number conserving symmetry, MPS algorithms will usually be accelerated if the quantum symmetry is explicitly takeing into account, namely using a U(1) symmetric MPS \cite{SinghVidal2010, GuoPoletti2019a}. We note that a non-symmetric matrix product states method has been applied to study time-bin quantum optics in \cite{Lubasch2018}, without using the number conserving property of the underlying optical quantum circuit. In particular, we highlight that we have developed a generic platform to study optical quantum circuits based on the U(1) symmetric MPS. It can be used to simulate different particle types including both the bosons and the fermions. Simulating an open quantum system can be done almost in parallel with the unitary case. The changes of the entanglement entropy as well as other physical observables during each depth of the evolution can be measured efficiently with this method.

The paper is organized as follows: In Sec.~\ref{sec:method}, we briefly introduce the MPS algorithms and how it is used to study optical quantum circuits. In Sec.~\ref{sec:result}, we show the entanglement entropy of the quantum state in a bosonic circuit is much larger and grows much faster than that in a fermionic circuit, and how the numerical complexity of simulating BosonSampling using MPS method is affected by losses. Finally we summarize in Sec.~\ref{sec:summary}.

\section{method} \label{sec:method}
In this section, we briefly show how to use MPS to simulate many-particle dynamics in various cases, namely bosons and fermions in an ideal environment as well as in a lossy environment.

A quantum many-body state living on a bosonic or fermionic optical lattice of size $M$ ($M$ modes) can be written as
\begin{align} \label{eq:generalwave}
\vert \psi \rangle = \sum_{n_1, n_2, \dots, n_M} c_{n_1, n_2, \dots, n_M} \vert n_1, n_2, \dots, n_M \rangle,
\end{align}
where $n_i$ labels the local Fock state on the $i$-th site of the lattice, $n_i$ is unbounded for bosons and $n_i=0,1$ for fermions. The coefficient $c_{n_1,\dots, n_M}$ is an $M$ dimensional tensor. Assuming the size of the local Hilbert space is $l$, then the tensor $c$ would contain $l^M$ complex numbers if quantum symmetry is not taken into account. MPS rewrites $c$ as a product of $3$ dimensional tensors:
\begin{align}\label{eq:generalmps}
c_{n_1, n_2, \dots, n_M} = \sum_{a_1, \dots, a_{M+1}} B_{a_1, a_2}^{n_1}B_{a_2, a_3}^{n_2}\dots B_{a_M, a_{M+1}}^{n_M}.
\end{align}
Here we have used $a_i$ to denote the auxiliary degree of freedom, with $a_1=a_{M+1}=1$ added for notational convenience. The largest size of $a_i$ is referred to as the bond dimension $D$, which essentially characterizes the complexity of MPS based algorithms.

\begin{figure} [htbp]
\includegraphics[width=\columnwidth]{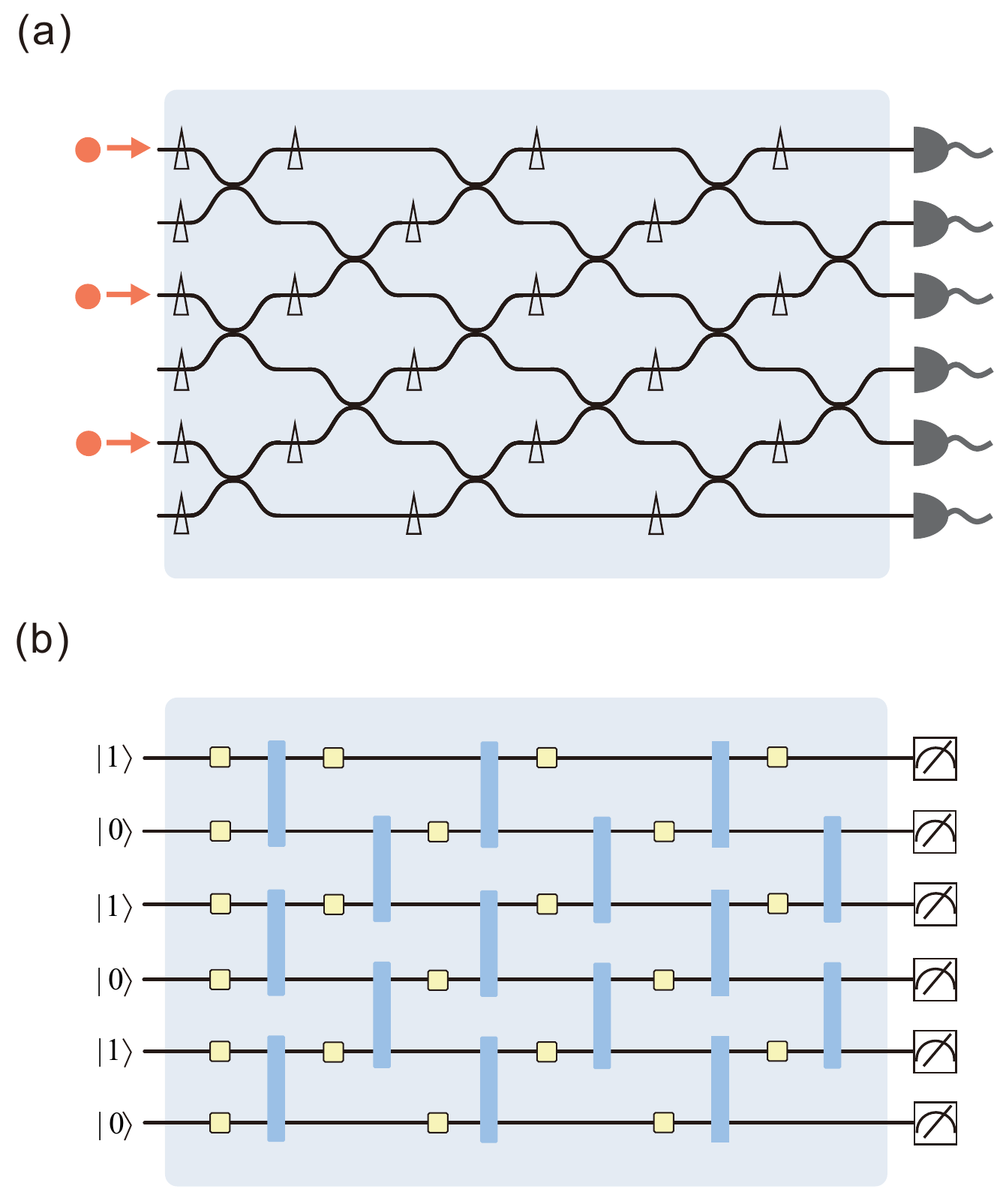}
\caption{ (a) Universal $M$-mode multiport interferometer (shown here for $M=6$) for BosonSampling \cite{Clements2016}. The triangle and the intersection of bent lines represent the phase shifter and beam splitter, respectively. Sequence of phase shifters and beam splitters acting on the optical quantum circuit. (b) Mapping from the optical quantum circuit to a quantum circuit consisting of one-body (yellow square, corresponding to the phase shifter) and two-body (blue rectangular, corresponding to the beam splitter) gate operations and acts on an MPS, in analogous to the t-MPS algorithm. A layer of two-body gate operations is counted as one depth.
 \label{fig:fig1} }
\end{figure}

MPS keeps invariant when inserting a pair of matrices $U$, $U^{-1}$ between a pair of sites $i$ and $i+1$ (such a pair of neighbouring sites is also referred to as a bond $i$). To fix this gauge degree of freedom, one can prepare the MPS in canonical forms \cite{Schollwock2011}. In our simulation we keep MPS in the right canonical form, which means that each tensor $B_{a_i, a_{i+1}}^{n_i}$ satisfies the condition
\begin{align} \label{eq:canonicalcondition}
\sum_{a_{i+1}, n_i} B_{a_i, a_{i+1}}^{n_i} \conj(B_{a_i', a_{i+1}}^{n_i}) = \delta_{a_i, a_i'},
\end{align}
where $\conj(a)$ means taking the element-wise conjugate of the tensor $a$.

Local unitary transformations on the quantum state in Eq.(\ref{eq:generalwave}) can be easily translated into operations on MPS. Here we exemplify this by showing how to apply one-body and nearest-neighbour two-body operations to MPS. The application of one-body operations $R_{n_i}^{n_i'}$ on site $i$ is simply
\begin{align}
B_{a_i, a_{i+1}}^{n_i'} = \sum_{n_i} R_{n_i}^{n_i'}B_{a_i, a_{i+1}}^{n_i}.
\end{align}
We can see that one-body operation will not break the right canonical condition in Eq.(\ref{eq:canonicalcondition}) as long as $R_{n_i}^{n_i'}$ is unitary.

For a two-body operation $O_{n_i, n_{i+1}}^{n_i', n_{i+1}'}$ acting on bond $i$, we first multiply $B_{a_i, a_{i+1}}^{n_i}$, $B_{a_{i+1}, a_{i+2}}^{n_{i+1}}$ and the singular vector $S_{a_i, a_i}$ between sites $i-1$ and $i$
\begin{align}
\Phi_{a_i, a_{i+2}}^{n_i, n_{i+1}} = \sum_{a_i, a_{i+1}}S_{a_i, a_i} B_{a_i, a_{i+1}}^{n_i}B_{a_{i+1}, a_{i+2}}^{n_{i+1}},
\end{align}
then we apply $O_{n_i, n_{i+1}}^{n_i', n_{i+1}'}$ to $\Phi$ and get
\begin{align}
\Phi_{a_i, a_{i+2}}^{n_i', n_{i+1}'} = \sum_{n_i,n_{i+1}} \Phi_{a_i, a_{i+2}}^{n_i, n_{i+1}} O_{n_i, n_{i+1}}^{n_i', n_{i+1}'},
\end{align}
after that we do a singular value decomposition (SVD) to $\Phi$
\begin{align}
\sum_{a_{i+1}'} U_{a_i, a_{i+1}'}^{n_i'} S_{a_{i+1}', a_{i+1}'} V_{a_{i+1}', a_{i+2}}^{n_{i+1}'} = \svd(\Phi_{(a_{i+2}, n_{i+1}')}^{(a_i, n_i')}),
\end{align}
where we have grouped the indexes $(a_{i+2}, n_{i+1}')$ and $(a_i, n_i')$ so that SVD is performed on a two-dimensional tensor. From the definition of SVD, the tensor $V$ satisfies the right canonical condition in Eq.(\ref{eq:canonicalcondition}) and the tensor $U$ satisfies the left canonical condition
\begin{align}
\sum_{a_i, n_i} U_{a_i, a_{i+1}}^{n_i} \conj(U_{a_i, a_{i+1}'}^{n_i}) = \delta_{a_{i+1}, a_{i+1}'}.
\end{align}
With these results, we can update $B_{a_{i+1}', a_{i+2}}^{n_{i+1}'}$ and $B_{a_{i}, a_{i+1}'}^{n_{i}'}$
\begin{align}
B_{a_{i+1}', a_{i+2}}^{n_{i+1}'} &= V_{a_{i+1}', a_{i+2}}^{n_{i+1}'}; \label{eq:B1} \\
B_{a_{i}, a_{i+1}'}^{n_{i}'} &= \sum_{a_i, a_{i+1}'} U_{a_i, a_{i+1}'}^{n_i'} S_{a_{i+1}', a_{i+1}'}/S_{a_{i}, a_{i}}. \label{eq:B2}
\end{align}
It is straightforward to verify that the new tensors $B_{a_{i+1}', a_{i+2}}^{n_{i+1}'}$ and $B_{a_{i}, a_{i+1}'}^{n_{i}'}$ produced in Eqs.(\ref{eq:B1}, \ref{eq:B2}) still satisfy the right canonical condition in Eq.(\ref{eq:canonicalcondition}). The division by very small singular values in Eq.(\ref{eq:B2}) could make this algorithm numerically unstable, which could be overcome by the algorithm in \cite{Hastings2009} with negligible additional effort. We note that since we are using a U(1) symmetric MPS, all the tensors mentioned above are assumed to be generic tensors which are protected by a global U(1) symmetry. Such symmetric tensors can be simply taken as a list of dense tensors which are labeled by good quantum numbers, whose corresponding tensor operations mimic those of  multi-dimensional dense tensors. The U(1) symmetric MPS leverages the fact that the total quantum number of the underlying quantum system is conserved, thus could often greatly reduce the memory requirements of MPS algorithms. For the implementation of generic tensor operations on U(1) symmetry protected tensors, one can refer to, for example \cite{SinghVidal2010}. The usage of U(1) symmetric MPS allows us to directly consider all the local degrees of freedom for bosonic particles, without the need to choose different truncations of the local Hilbert space until the result converges, as is done in \cite{Lubasch2018}.

In BosonSampling, only phase shifters and beam splitters are used. A phase shifter is a one-body operation, and a beam splitter is a two-body operation. Moreover, as in Fig.~\ref{fig:fig1}, the beam splitter acts only on nearest-neighbouring sites. Therefore the above operations would suffice for BosonSampling. We also note that the successive application of phase shifters and beam splitters to  MPS would be analogous to the time-dependent matrix product states (t-MPS) algorithm \cite{Vidal2003, DaleyVidal2004, WhiteFeiguin2005, Schollwock2011}.

\subsection{Bosonic circuit}
One important property of a optical bosonic circuit made of phase shifters and beam splitters is that the total number of particles in the system is conserved. Namely if the input quantum state has a fixed total number of particles $N$, then during the evolution, $N$ is preserved. This means that we only need to consider the coefficients $c_{n_1, n_2, \dots, n_M}$ which satisfy
\begin{align} \label{eq:numberconserve}
\sum_{i=1}^M n_i = N.
\end{align}
Without loss of generality, the initial state $\vert \phi \rangle$ is chosen so that the first $N$ modes have one photon as inputs, while for the rest there is no input photon. This state could be written as
\begin{align}
\vert \phi \rangle = \vert 1 \rangle^N \otimes \vert 0\rangle^{M-N}.
\end{align}

To this end, we note that in the literature the phase shifter and beam splitter are often written as linear mappings in the operator space, namely mapping from one quantum operator to another. To use them with MPS, one need to rewrite them as quantum operators which map one quantum state to another. A phase shifter with a rotation angle $\phi$ can straightforwardly be written as
\begin{align}
\phaseshifter^b(\phi)\vert n \rangle = e^{-2 \pi \im n\phi} \vert n \rangle.
\end{align}
For a beam splitter with an angle $\theta$, it is usually written in the operator space as
\begin{align}
\beamsplitter^b(\theta)\left(\begin{array}{c}
\adop \\
\bdop
\end{array}\right) = \left(\begin{array}{cc}
\cos(2\pi\theta) & -\sin(2\pi\theta) \\
\sin(2\pi\theta) & \cos(2\pi\theta)
\end{array}\right) \left(\begin{array}{c}
\adop \\
\bdop
\end{array}\right)\end{align}
where $\adop$ and $\bdop$ are bosonic creation operators. Written as an operator on a quantum state $\vert m, n \rangle$, it would be
\begin{align} \label{eq:bosonbeamsplitter}
\beamsplitter^b(\theta)\vert m, n \rangle =& \beamsplitter^b(\theta) \frac{(\adop)^m}{\sqrt{m!}}\frac{(\bdop)^n}{\sqrt{n!}} \vert 0, 0\rangle \nonumber \\
=& \frac{\left(\cos(2\pi\theta)\adop-\sin(2\pi\theta)\bdop\right)^m}{\sqrt{m!}} \nonumber \\
&\frac{\left(\sin(2\pi\theta)\adop+\cos(2\pi\theta)\bdop\right)^n}{\sqrt{n!}}\vert 0, 0\rangle
\end{align}
For example, for an input state $\vert 1, 1\rangle$, Eq.(\ref{eq:bosonbeamsplitter}) will be
\begin{align}
\beamsplitter^b(\theta)\vert 1, 1 \rangle =& \sqrt{2}\cos(2\pi\theta)\sin(2\pi\theta)\vert 2, 0\rangle \nonumber \\
 &- \sqrt{2}\cos(2\pi\theta)\sin(2\pi\theta)\vert 0, 2\rangle \nonumber \\
 &+ \left(\cos^2(2\pi\theta) - \sin^2(2\pi\theta)\right) \vert 1, 1\rangle.
\end{align}
To ensure that the optical quantum circuit corresponds to a Haar random unitary matrix, we use the approach in \cite{RussellLaing2017}, namely each $\phi$ is generated according to a uniform distribution, while each $\theta$ is generated according to a specific distribution which is dependent on the depth and the position.

\subsection{Fermionic circuit}
In comparison with BosonSampling, sampling fermions (we call it FermionSampling in this paper) is related to calculating determinants of matrices, for which there exists efficient classical algorithms. Nevertheless, it would still be insightful to see whether FermionSampling is easy to simulate or not with MPS based algorithms. The simulation of FermionSampling is done by mapping the fermionic circuit to a one-dimensional spin chain by Jordan-Wigner transformation \cite{JordanWigner1928, LiebMattis1961}. As a result, the beam splitter for the fermionic operator $\beamsplitter^f$, in the zero- and one-particle sections, would remain the same as Eq.(\ref{eq:bosonbeamsplitter}), while in the two-particle section, it is trivially
\begin{align}
\beamsplitter^f(\theta) \vert 1, 1\rangle = \vert 1, 1\rangle,
\end{align}
due to the fermionic anti-commutation relation.

\subsection{Lossy bosonic circuit}
We also consider bosonic circuit in a lossy environment \cite{Garcia-PatronShchesnovich2017, OszmaniecBrod2018, RenemaGarcia-Patron2019}. In particular, we consider the simple case studied in \cite{Garcia-PatronShchesnovich2017}, where it is assumed that photons are injected into an optical circuit with uniform-loss channels of transmission ($i$.$e$. affecting all modes equally), and only the loss in the circuit is considered. In this simple and special case, the lossy linear optical circuit can be modeled by a lossless circuit, but with input to be a mixed state instead of a pure state. In this case, the quantum state is a mixture of different total number of particles. Nevertheless, by using the method in \cite{GuoPoletti2019a}, we could still benefit from a U(1) symmetric MPS. Such an input state can be written as
\begin{align}\label{eq:rho_density}
\rho = \sigma^N \otimes \vert 0\rangle\langle 0\vert^{M-N},
\end{align}
where each $\sigma$ is a local density operator. Here $N$ means the largest possible total number of particles instead. To deal with this kind of input state with MPS, one could reshape $\rho$ into a long vector which we denote as $\vert\rho\rangle\rangle$
\begin{align}
\vert \rho\rangle\rangle = \vert \sigma\rangle\rangle^N \otimes \vert 00\rangle\rangle^{M-N}.
\end{align}
Here $\vert \sigma\rangle\rangle$ and $\vert 00\rangle\rangle$ denote vectors resulting from squashing of the local density operators $\sigma$ and $\vert 0\rangle\langle 0\vert$ in Eq.(\ref{eq:rho_density}). The local Hilbert space would then be enlarged to $(N+1)\times(N+1)$. Correspondingly, any unitary gate $U$ operating on the corresponding unitary system, will become
\begin{align}
U^{o} = \kron (U, \conj(U)).
\end{align}
Here $\kron(a, b)$ means the kronecker product of two tensors $a$ and $b$.

We note the usage of U(1) symmetric MPS here allows us to choose much larger bond dimension $D$. In our simulations of the lossy cases, we have made use of this properties, which allows us to study an open system with $D$ larger than $10000$. In comparison, with a non-symmetric MPS, this case is almost intractable with a personal computer since the size of the local Hilbert space $l$ will already be $49$ for $N=6$ (A two-body operation would require to do an SVD on a dense matrix of size $Dl\times Dl$).

For simplicity, in this work we focus on a special type of initial local mixed state $\sigma$, which is
\begin{align}
\sigma = \bar{n}\vert 1\rangle\langle 1\vert + (1-\bar{n})\vert 0\rangle\langle 0\vert,
\end{align}
$\bar{n}$ is the average filling of bosons which satisfies $0\leq \bar{n}\leq 1$. However, more general intial mixed state, as well as more general losses could also be studied with our setup.

\section{result} \label{sec:result}

\begin{figure}[htbp]
\includegraphics[width=\columnwidth]{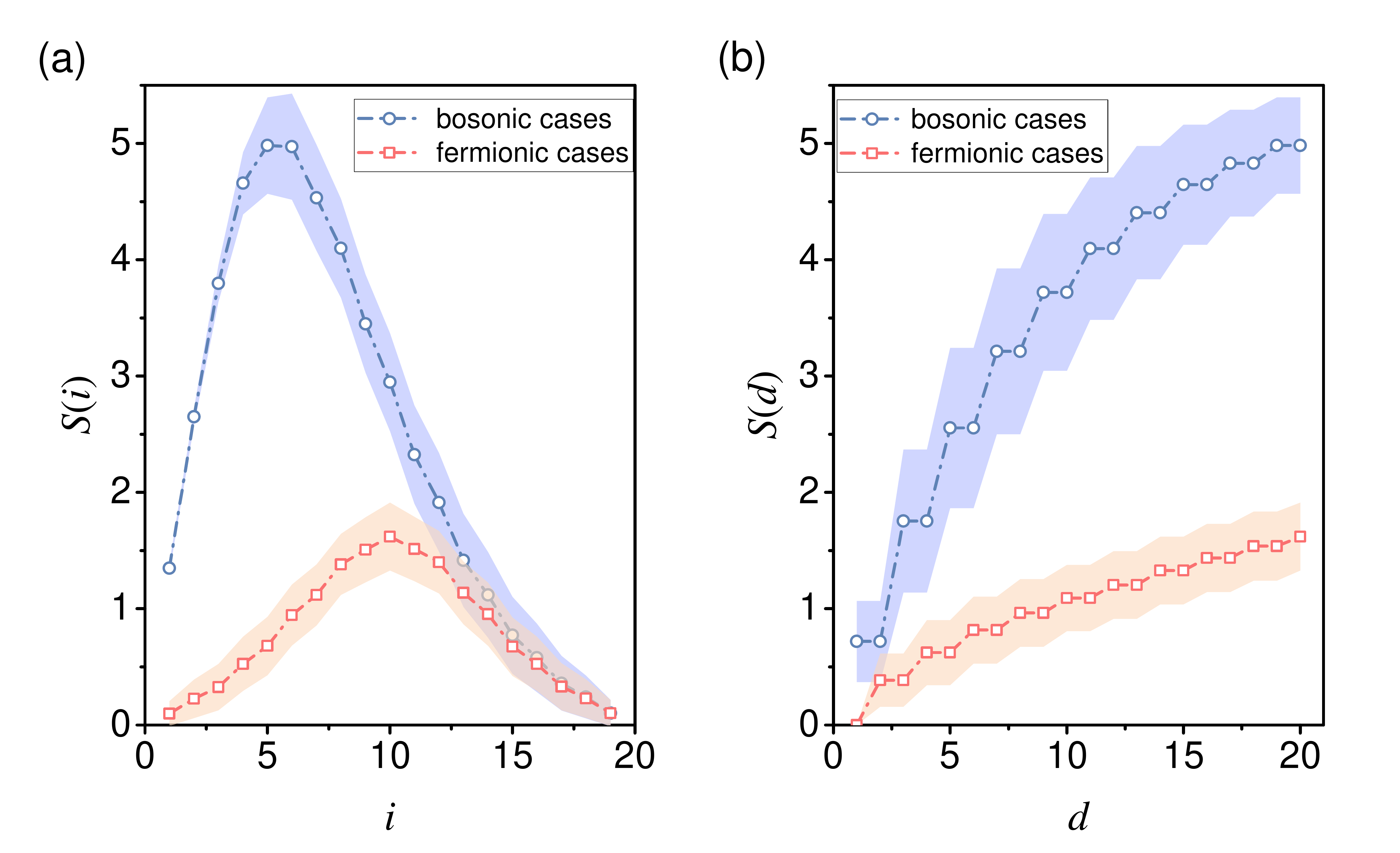}
\caption{ (a) Entanglement entropy  $\entropy$ of the final quantum state as a function of the bond index $i$ which corresponds to the place of the bipartition. The depth $d= 20$. (b) Entanglement entropy $\entropy$ grows as a function of depth $d$, corresponding to the bond for which $\entropy$ grows fastest. The system size used for both figures is $M=20$ and $N=10$, and the results are averaged over $10^3$ random generated optical quantum circuits. In both figures, the blue line with circle represents the bosonic case while the red line with square represents the fermionic case. The painted area represents the standard deviation.
\label{fig:fig2} }
\end{figure}

An important advantage of MPS is that one could explicitly monitor the growth of entanglement entropy $\entropy$ during the dynamics, which is defined as
\begin{align}\label{eq:entropy}
\entropy = - \tr\left[\rho_A\log(\rho_A)\right], \rho_A = \tr_B\rho.
\end{align}
Here the subscripts $A$ and $B$ mean two subsystems resulting from a bipartition of the system, and the density operator $\rho=\vert\phi\rangle\langle \phi\vert$ with $\vert\phi\rangle$ being the wave function of the system. In Fig.~\ref{fig:fig2}, we randomly generate $10^3$ optical quantum circuits corresponding to Haar unitary matrices for bosons and fermions with $M=20$ and $N=10$, and evolve them to a maximum depth $d_{\max}=M=20$. For all the simulations done in this work, we have chosen a maximum bond dimension $D=15000$ to ensure that it would neven be reached, and an SVD truncation threshold $t=10^{-8}$. In Fig.~\ref{fig:fig2}(a) we plot the distribution of entropy $\entropy$ of the final quantum state as a function of the bond index $i$ corresponding to the bipartition position, for both the bosonic and the fermionic cases. We can see that in the bosonic case, the distribution is unbalanced (not reflection-symmetric around the central bond) which is a remnant of the unbalanced initial state, while in the fermionic case the distribution is balanced irrespective of the initial state. In in Fig.~\ref{fig:fig2}(b), we plot the growth of entropy $\entropy$ as a function of the evolution depth $d$. The entanglement entropy $\entropy$ of the quantum system strongly affects our ability to simulate it with MPS algorithms \cite{Schollwock2011}. From Fig.~\ref{fig:fig2}(a) and Fig.~\ref{fig:fig2}(b), we could see clearly that the entanglement entropy $\entropy$ of the quantum state in a bosonic circuit is much larger and grow much faster than that in a fermionic circuit, which means that simulating the multi-particle evolution in a bosonic circuit is more difficult than that in a fermionic circuit using MPS. However, we must note that although the growth of entanglement entropy is the key factor can directly affect the efficiency of MPS \cite{Schollwock2011}, it is not equivalent to the classical hardness of BosonSampling, as there are cases with entanglement at the output that are classically simulatable \cite{Tichy2014}.


In fact, it is very simple and natural to measure $\entropy$ during the dynamical process in the optical quantum circuit in Fig.~\ref{fig:fig1}. The singular values are stored alongside with the MPS whenever a two-body gate operation is performed, thus allowing to compute the entropy with almost no additional effort. We can simulate the evolution of photons in a BosonSampling circuit with $M=26$, $N=13$ and $d_{\max}=26$ in only about $5500$s on a personal computer with $2$ cores of 2.8 GHz frequency and $16$ Gb memory, where we have measured the entropy $\entropy$ of each depth. We note that if we measure the entropy by first reconstructing the quantum state at each depth, we need to spend a huge amount of time to calculate all of the amplitudes of a quantum state. Moreover, it would require about $87$ Gb memory on a classical computer if one stores all these amplitudes exactly.  Concretely, a pure python program using the Ryser's algorithm is able to compute the permanent of a matrix of size $13$ in $0.09$s. Taking into considerations the huge number of matrix permanents to be computed to reconstruct the quantum state, MPS will be $10^{6}$ times faster than the Ryser's algorithm. We note that here we only compare our method to a brute-force simulation that uses Ryser's algorithm, but more efficient alternatives may exist.

\begin{figure}[htbp]
\includegraphics[width=\columnwidth]{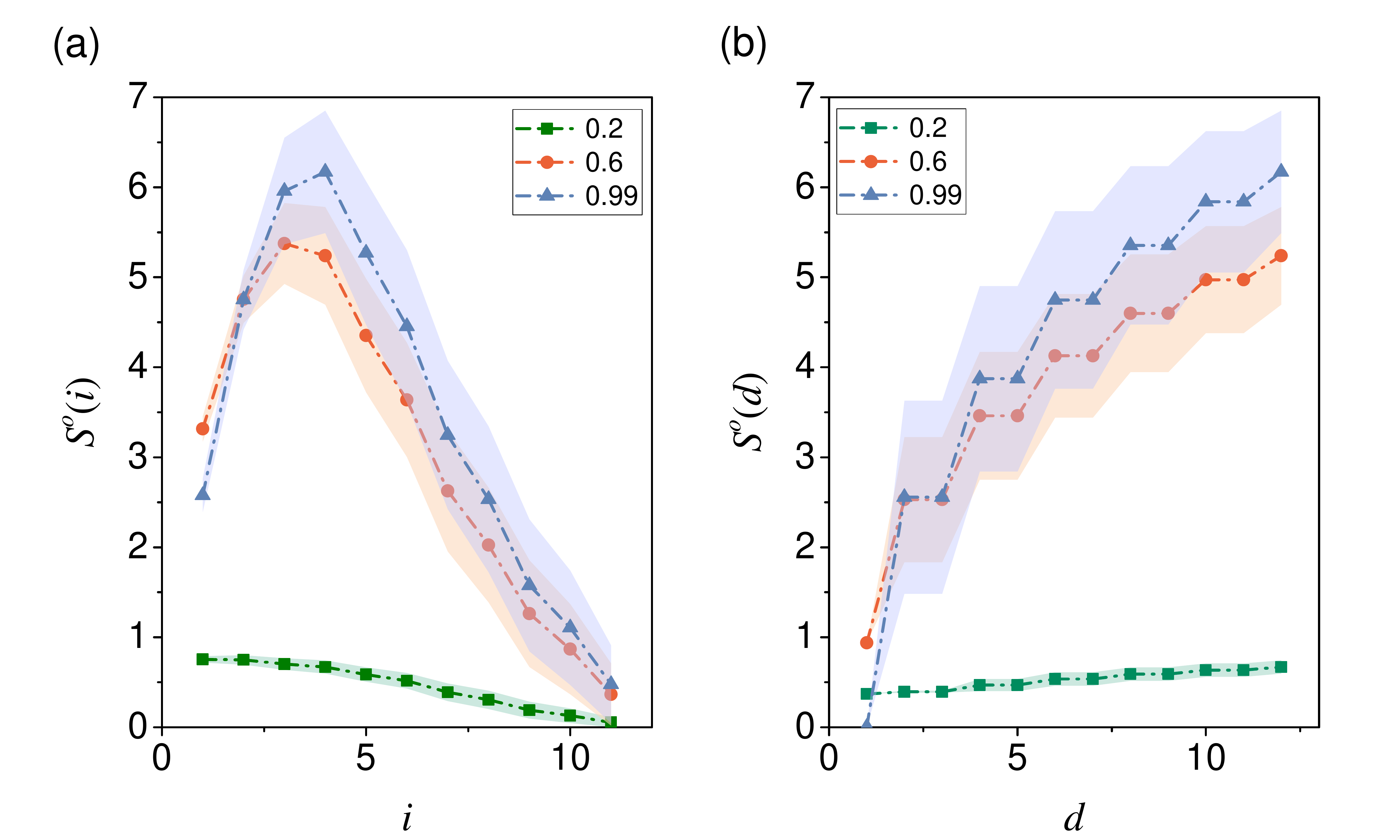}
\caption{ (a) OSEE $\entropy^o$ of the final quantum state as a function of the bond index $i$ which corresponds to the place of the bipartition. (b) OESS $\entropy^o$ grows as a function of depth $d$, corresponding to the bond for which $\entropy^o$ grows fastest. The system size used for both figures is $M=12$, and the results are averaged over $10^3$ random generated optical quantum circuits. In both figures, the blue line with triangle, the red line with circle and the green line with square represent $\bar{n}=0.99, 0.6, 0.2$, respectively. The painted area represents the standard deviation.
\label{fig:fig3} }
\end{figure}

We also apply MPS to study BosonSampling in presence of losses, which is called lossy BosonSampling \cite{Aaronson2016}. The hardness of MPS used to represent a density operator for an open quantum system can be evaluated by the operator space entanglement entropy (OSEE), which we denote as $\entropy^o$ in comparison to the unitary case in Eq.(\ref{eq:entropy}). Mathematically, OSEE is defined as \cite{Prosen2008a}
\begin{align}
\entropy^{o} = -\tr_A\left[R\log(R)\right], \space R = \langle\langle \rho \vert \vert \rho \rangle\rangle^{-1}\tr_B \vert\rho\rangle\rangle\langle\langle\rho\vert.
\end{align}
In Fig.~\ref{fig:fig3}(a), we plot the distribution of $\entropy^o$ as a funciton of bond index $i$, while in Fig.~\ref{fig:fig3}(b) we plot the growth of $\entropy^o$ as a function of the evolution depth $d$, for $\bar{n}=0.99, 0.6, 0.2$ and for bosonic particles. We can see that as the average occupation $\bar{n}$ goes down, $\entropy^o$ grows much slower, which is in correspondence with the theoretical results in Ref. \cite{Aaronson2016}.

Here we point out that $\entropy^o$ is essentially a different concept than $\entropy$ and does not represent the entanglement of a density operator \cite{Prosen2008a}. It just characterizes the hardness of an density operator written as an MPS. Thus, we cannot study the difference between ideal BosonSampling and lossy BosonSampling by directly comparing the entanglement entropy $S$ and the OSEE $\entropy^o$. Moreover, the entropy $\entropy$ and OSEE $\entropy^o$ only indicate the difficulty of a problem when simulated with MPS algorithms, which are not necessarily equivalent to the complexity of the underlying problem.

\begin{figure}[tbp]
\includegraphics[width=0.9\columnwidth]{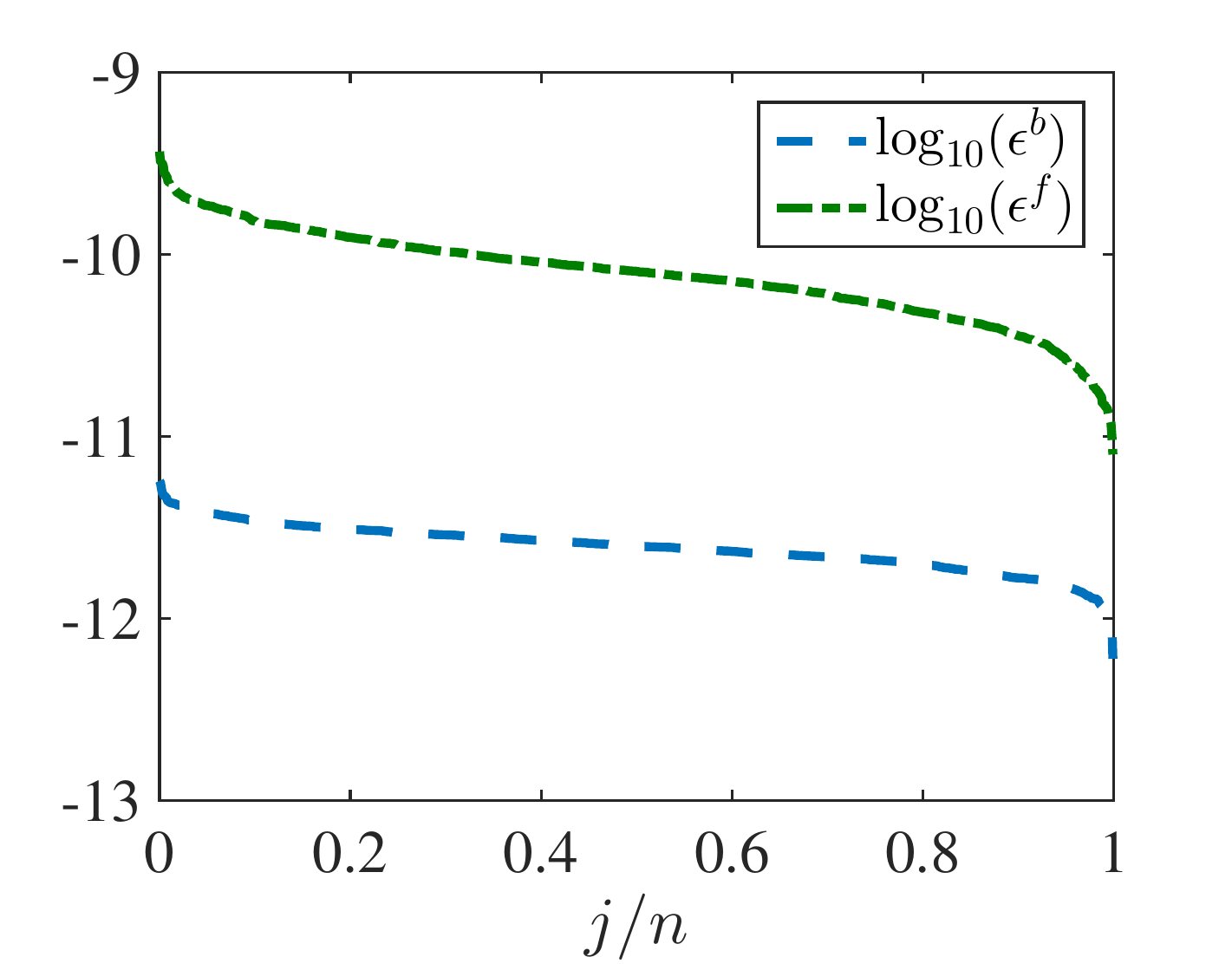}
\caption{ Sampling quality of MPS method with an SVD truncation threshold $t=10^{-8}$. In the bosonic case (blue dashed line), the error is denoted by the distances between all the samples of the output probability computed with MPS and that computed with Ryser's algorithm. In the fermionic case (green dot-dashed line), the error is denoted by the distances between all the samples computed with MPS and that computed with the python built-in $\det$ function. The system size for both cases are $M=20, N=10$. The total number of randomly generate circuits $n=10^3$ and the total number of ranomly chosen basis $K=10^4$.
 \label{fig:fig4} }
\end{figure}

To this end, we note that MPS is an approximate representation of the quantum state, in that each time a two-body gate operation is performed, an SVD is done and the singular values under the threshold $t$ are thrown away. Therefore it is possible that such small errors during the evolution will accumulate after many steps of evolution. The number of two-body gate operations in the circuit is approximately $M^2/2$. Thus in the worst case the error $\epsilon$ for the output probability calculated by MPS would be $\epsilon \approx M^2 t/2$ (number of two-body operations times the truncation threshold for each two-body operation). For the case $M=20$ and $N=10$, the value of $\epsilon$ would be around $2\times 10^{-6}$ at most. To test the overall quality of the MPS method, we generate $10^3$ random optical quantum circuits. For each circuit, we calculate $K$ output probabilities according to $K$ randomly chosen basis, where we have used $K=10^4$ in our simulation. Each output probability computed with MPS is denoted as $\perm^{{\rm MPS}}_{j, k}$ in the bosonic case and $\detm^{{\rm MPS}}_{j, k}$ in the fermionic case, where $j$ labels the $j$-th random optical circuit and $k$ labels the $k$-th random basis.  In Fig.~\ref{fig:fig4}, We compare $\perm^{{\rm MPS}}_{j, k}$ with the samples from a brute force application of the Ryser's algorithm, denoted as $\perm^{{\rm Ryser}}_{j, k}$ in the bosonic case, and compare $\detm^{{\rm MPS}}_{j, k}$ with $\detm_{j, k}$ computed with the python built-in $\det$ function in the fermionic case. Each point on Fig.~\ref{fig:fig4} corresponds to
\begin{align}
\epsilon^b_j &=\sqrt{\sum_{k=1}^{K}\left(\perm^{{\rm MPS}}_{j, k} - \perm^{{\rm Ryser}}_{j, k}\right)^2} \\
\epsilon^f_j  &=\sqrt{\sum_{k=1}^{K} \left(\detm^{{\rm MPS}}_{j, k} - \detm_{j, k}\right)^2}.
\end{align}
We can see that the numerical difference between our results with the standard algorithm reaches smaller than $10^{-9}$, indicating a remarkable accuracy of our approach. We could also increase this precision by a lower $t$, which would also mean that more resources would be required.

\section{Conclusion} \label{sec:summary}

In summary, we have used a U(1) symmetric MPS to study the dynamical evolution of photons in the optical quantum circuit. We monitor the entanglement entropy growth during the dynamical process, and show that the entropy of a bosonic circuit grows much faster than that of a fermionic circuit. We have also studied lossy BosonSampling, showing that as the loss increases, the operator space entanglement entropy grows slower, which means that classically simulating an optical circuit with larger loss using MPS could be much easier than an ideal one. MPS provides a generic platform to study universal multi-particle quantum walks. More investigations into interesting quantum phenomena in other types of multi-particle quantum walks could be carried out in future works. Other BosonSampling protocols that could significantly reduce the physical resource requirements, and validation methods based on high-order correlation measurements, such as the work in Ref.~\cite{Walschaers2016,Huang2017}, could also be further investigated.

\begin{acknowledgments}
We thank Dario Poletti and Zu-En Su for helpful discussions. C.G. acknowledges support from National Natural Science Foundation of China under Grants No. 11504430 and No. 11805279. H.-L. Huang acknowledges support from the Open Research Fund from State Key Laboratory of High Performance Computing of China (HPCL) (No. 201901-01)
\end{acknowledgments}


\begin{thebibliography}{99}
\bibitem{pan2012} J.-W. Pan, Z.-B. Chen, C.-Y. Lu, H.Weinfurter, A. Zeilinger, and M. \.Zukowski, Rev. Mod. Phys.  {\bf 84}, 777 (2012).
\bibitem{sihui2012} S.-H. Tan and P. P. Rohde, arXiv:1805.11827 (2018).
\bibitem{kok2007} P. Kok, W. J. Munro, K. Nemoto, T. C. Ralph, J. P. Dowling, and G. J. Milburn, Rev. Mod. Phys. {\bf 79}, 135 (2007).
\bibitem{lo2012} H. K. Lo, M. Curty, and B. Qi,  Phys. Rev. Lett. {\bf 108(13)}, 130503 (2012).
\bibitem{Guzik2012} A. Aspuru-Guzik and P. Walther, Nat. Phys. {\bf 8}, 285 (2012).
\bibitem{Giovannetti2011} V. Giovannetti, S. Lloyd, and L. Maccone, Nat. Photonics {\bf 5}, 222 (2011).
\bibitem{hong1987} C. K. Hong, Z. Y. Ou, and L. Mandel, Phys. Rev. Lett. {\bf 59}, 2044 (1987).
\bibitem{Lund2017} A. P. Lund, M. J. Bremner, and T. C. Ralph, npj Quantum Inf., {\bf 3(1)}, 15 (2017).
\bibitem{Aaronson2013} S. Aaronson and A. Arkhipov, Theory Comput. {\bf 9}, 143 (2013).
\bibitem{Valiant1979} L. G. Valiant, Theoretical computer science {\bf 8}, 189 (1979).
\bibitem{Harrow2017} A. W. Harrow, and A. Montanaro, Nature, {\bf 549(7671)}, 203 (2017).
\bibitem{Broome2013} M. A. Broome, A. Fedrizzi, S. Rahimi-Keshari, J. Dove, S. Aaronson, T. C. Ralph, and A. G. White, Science {\bf 339}, 794 (2013).
\bibitem{Tillmann2013} M. Tillmann, B. Daki\'c, R. Heilmann, S. Nolte, A. Szameit, and P. Walther, Nat. Photonics {\bf 7}, 540 (2013).
\bibitem{Crespi2013}  A. Crespi, R. Osellame, R. Ramponi, D. J. Brod, E. F. Galv\~ao, N. Spagnolo, C. Vitelli, E. Maiorino, P. Mataloni, and F. Sciarrino, Nat. Photonics 7, 545 (2013).
\bibitem{Spring2013} J. B. Spring $et$ $al.$, Science {\bf 339}, 798 (2013).
\bibitem{Spagnolo2014} N. Spagnolo $et$ $al$., Nat. Photonics {\bf 8}, 615 (2014).
\bibitem{Carolan2014}  J. Carolan $et$ $al$., Nat. Photonics {\bf 8}, 621 (2014).
\bibitem{Carolan2015} J. Carolan $et$ $al$., Science {\bf 349}, 711 (2015).
\bibitem{he2017} Y. He $et$ $al$., Phys. Rev. Lett. {\bf 118}, 190501 (2017).
\bibitem{wang2017} H. Wang  $et$ $al$., Nat. Photonics {\bf 11}, 361-365 (2017).
\bibitem{Loredo2017}  J. C. Loredo, M. A. Broome, P. Hilaire, O. Gazzano, I. Sagnes, A. Lemaitre, M. P. Almeida, P. Senellart, and A. G. White, Phys. Rev. Lett. {\bf 118}, 130503 (2017).
\bibitem{wang2018} H. Wang  $et$ $al$., Phys. Rev. Lett. {\bf 120}, 230502 (2018).
\bibitem{Urbina2014} J.-D. Urbina, J. Kuipers, Q. Hummel, and K. Richter, arXiv:1409.1558 (2014).
\bibitem{Tamma2015}V. Tamma and S. Laibacher, Phys. Rev. Lett. {\bf 114}, 243601 (2015).
\bibitem{Laibacher2018} S. Laibacher and V. Tamma, arXiv:1801.03832 (2018).
\bibitem{Shchesnovich2016} V. S. Shchesnovich, Phys. Rev. Lett. {\bf 116}, 123601 (2016).
\bibitem{Walschaers2016} M. Walschaers, J. Kuipers, J.-D. Urbina, K. Mayer, M. C. Tichy, K. Richter, and A. Buchleitner, New J. Phys. {\bf 18}, 032001 (2016).

\bibitem{Huang2017} H.-L. Huang, H.-S. Zhong, T. Li, F.-G. Li, X.-Q. Fu, S. Zhang, X. Wang and W.-S. Bao, Sci. Rep., {\bf 7(1)}, 15265 (2017).

\bibitem{Neville2017} A. Neville, C. Sparrow, R. Cliord, E. Johnston, P. M. Birchall, A. Montanaro, and A. Laing, Nat. Phys. {\bf 13}, 1153 (2017).
\bibitem{WuYang2018} J. Wu, Y. Liu, B. Zhang, X. Jin, Y. Wang, H. Wang, and  X. Yang, Natl. Sci Rev., {\bf 5(5)}, 715-720 (2018).
 \bibitem{Ryser1963} H. J. Ryser, in $Combinatorial$ $Mathematics$ (The Mathematical Association of America, 1963), p. 154.
\bibitem{CliffordClifford2017} Peter Clifford and Rapha\"el Clifford, arXiv:1706.01260 (2017).

\bibitem{Perez-GarciaCirac2007} D. Perez-Garcia, F. Verstraete, M. M. Wolf, J. I. Cirac, Quantum Inf. Comput. {\bf 7}, 401 (2007)
\bibitem{Schollwock2011} U. Schollwo\"ck, Ann. Phys. {\bf 326}, 96 (2011).
\bibitem{Tichy2014} M. C. Tichy, K. Mayer, A. Buchleitner, and K. M{\o}lmer, Phys. Rev. Lett., {\bf 113(2)}, 020502 (2014).

\bibitem{White1992} S. R. White, Phys. Rev. Lett., {\bf 69}, 2863 (1992)
\bibitem{White1993} S. R. White, Phys. Rev. B, {\bf 48}, 10345 (1993)
\bibitem{DelayVidal2004} A. J. Daley, C. Kollath, U. Schollw\"ock, G. Vidal, J. Stat. Mech: Theor. Exp., (2004), {\bf P}04005
\bibitem{GuoPoletti2018} C. Guo, I. de Vega, U. Schollw\"ock, D. Poletti, Phys. Rev. A, {\bf 97}, 053610 (2018)
\bibitem{BernierPoletti2018} J. S. Bernier, R. Tan, L. Bonnes, C. Guo, D. Poletti, C. Kollath, Phys. Rev. Lett., {\bf 120}, 020401 (2018)
\bibitem{XuPoletti2018} X. Xu, J. Thingna, C. Guo, D. Poletti, arXiv:1806.09181 (2018)

\bibitem{Hastings2007} M. B. Hastings, J. Stat. Mech: Theory Exp. (2007) {\bf P}08024.

\bibitem{SinghVidal2010} Sukhwinder Singh, Robert N. C. Pfeifer, and Guifre Vidal, Phys. Rev. B {\bf 83}, 115125 (2011).

\bibitem{GuoPoletti2019a} Chu Guo, Dario Poletti, arXiv:1905.01609 (2019).

\bibitem{Lubasch2018} Michael Lubasch, Antonio A. Valido, Jelmer J. Renema, W. Steven Kolthammer, Dieter Jaksch, M. S. Kim, Ian Walmsley, and Ra\'ul Garc\'ia-Patr\'on, Phys. Rev. A, {\bf 97}, 062304 (2018).

\bibitem{Clements2016} W. R. Clements, P. C. Humphreys, B. J. Metcalf, W. S. Kolthammer, and I. A. Walsmley, Optica {\bf 3}, 1460 (2016).


\bibitem{Hastings2009} M. Hastings,  J. Math. Phys. {\bf 50}, 095207 (2009).
\bibitem{Vidal2003} G. Vidal, Phys. Rev. Lett. {\bf 91}, 147902 (2003).
\bibitem{DaleyVidal2004} A. J. Daley, C. Kollath, U. U. Schollw\"ock, and G. Vidal, J. Stat. Mech. (2004) P04005.
\bibitem{WhiteFeiguin2005} S. R. White and A. E. Feiguin, Phys. Rev. Lett. {\bf 93}, 076401 (2004).

\bibitem{RussellLaing2017} Nicholas J Russell, Levon Chakhmakhchyan, Jeremy L O'Brien and Anthony Laing, New J. Phys. {\bf 19}, 033007 (2017).


\bibitem{JordanWigner1928} P. Jordan, E. Wigner, Z. Physik {\bf 47}, 631 (1928).
\bibitem{LiebMattis1961} E. Lieb, T. Schultz, D. Mattis, Ann. of Phys. {\bf 16}, 407 (1961).

\bibitem{Garcia-PatronShchesnovich2017} Ra\'ul Garc\'ia-Patr\'on, Jelmer J. Renema, Valery Shchesnovich, arXiv:1712.10037 (2017).

\bibitem{OszmaniecBrod2018} Micha{\l} Oszmaniec, Daniel J. Brod, New J. Phys. {\bf 20}, 092002 (2018).

\bibitem{RenemaGarcia-Patron2019} J.J. Renema, V. Shchesnovich, R. Garcia-Patron, arXiv:1809.01953 (2019).


\bibitem{Aaronson2016} S. Aaronson and D. J. Brod, Phys. Rev. A {\bf 93}, 012335 (2016).
\bibitem{Prosen2008a} T. Prosen, M. Znidaric, J. Stat. Mech. (2009) {\bf P}02035.

\end{thebibliography}
\end{document}